\pdfoutput=0

\documentclass[aps,prl,notitlepage,groupedaddress]{revtex4-1}
\usepackage{graphicx}
\usepackage{dcolumn}
\usepackage{bm}
\usepackage{epstopdf}
\usepackage{natbib}
\usepackage{amsmath}
\usepackage{amsfonts}
\usepackage{comment}

\newcommand\Rey{\mbox{\textit{Re}}}  

\newsavebox{\astrutbox}
\sbox{\astrutbox}{\rule[-5pt]{0pt}{20pt}}

\newcommand\eg{e.g.\ }

\newcommand{\ud}{\;\mathrm{d}} 
\newcommand{\ppartial}{\partial^2} 
\newcommand{\beq}{\begin{equation}} 
\newcommand{\eeq}{\end{equation}}
\newcommand{\beqa}{\begin{eqnarray}} 
\newcommand{\eeqa}{\end{eqnarray}}

\newcommand{\ie}{i.e. }
\newcommand{\nablabf}{\boldsymbol \nabla}

\newcommand{\ebf}{\boldsymbol e}

\newcommand{\ez}{\ebf_z}
\newcommand{\ubf}{\boldsymbol u}
\newcommand{\rbf}{\boldsymbol r}
\newcommand{\rperpbf}{\boldsymbol r_\perp}
\newcommand{\tbf}{\boldsymbol t}
\newcommand{\nbf}{\boldsymbol n}
\newcommand{\sech}{\,\text{sech}\,}

\begin{document}
\title{Slow flow in channels with porous walls}
\author{Kaare H Jensen}
\affiliation{Department of Physics, Technical University of Denmark, Kgs. Lyngby, Denmark}
\altaffiliation{Department of Organismic and Evolutionary Biology, Harvard University, Cambridge, MA, USA}
\altaffiliation{Materials Research Science and Engineering Center, Harvard University, Cambridge, MA, USA}
  \email{jensen@fas.harvard.edu}

\date{\today}
\begin{abstract}
We consider the slow 
flow of a viscous incompressible liquid  in a channel of constant but arbitrary cross section shape, driven by non-uniform suction or injection through the porous channel walls. 
A similarity transformation reduces the Navier-Stokes equations to a set of coupled equations for the velocity potential in two dimensions. When the channel aspect ratio and Reynolds number are both small, the problem reduces to solving the biharmonic equation with constant forcing in two dimensions. With the relevant boundary conditions, determining the velocity field in a porous channels is thus  equivalent to solving for the vertical displacement of a simply suspended thin plate under uniform load. This allows us to provide analytic solutions for flow in porous channels whose cross-section is e.g. a rectangle or an equilateral triangle, and provides a general framework for the extension of Berman flow (Journal of Applied Physics \textbf{24}(9), p. 1232, 1953) to three dimensions.
%
%
%
\end{abstract}

\maketitle

\section{Introduction}
Channel flows -- liquid flows confined within a closed conduit with no free surfaces -- are ubiquitous. In animals \citep{LaBarbera1990} and plants \citep{Holbrook2005} they serve as the building blocks of vascular systems, distributing energy to where it is needed and allowing distal parts of the organism to communicate.
When constructed by humans, one of the major functions of channels is to transport liquids or gasses,  e.g. water (irrigation and urban water systems) and energy (oil or natural gas) from sites of production to the consumer or industry.

In some cases, the channels have solid walls which are impermeable to the liquid flowing inside. In other cases, the channels have porous walls which allow the liquid to flow across the wall and thus modify the axial flow. Both are important.  The first class of flow has been studied in great detail, and analytical solutions are known in a few, but important, cases \citep{Batchelor2000}. The latter class has received much less attention, although it is equally important. The effect of porous walls is especially important in the study of biological flows due to the presence of permeable cell walls \citep{Holbrook2005}  and in industrial filtration applications \citep{Nielsen2012}.

Some analytic solutions of the flow in porous walled channels are known, primarily due to a similarity technique first used in this context by \citet{Berman1953}. Berman's method is closely related to those commonly used in boundary layer theory \citep{Schlichting2000} and allows for the solution of steady flows in geometries with symmetries which makes the problem two-dimensional. By demanding that the solution be of similarity form the Navier-Stokes equation is reduced to a single non-linear third-order differential equation for the velocity potential in one space dimension. The flow between parallel plates \citep{Berman1953} and in a cylindrical \citep{Yuan1956a} and annular tube \citep{Berman1958} have been analyzed in this way. Time dependent flows, flows at high Reynolds numbers and questions of uniqueness and stability of these flows have since been address by a large number of workers, see e.g. \cite{Cox1991}.

In this paper, we extend Berman's method to three dimensional similarity flows, and derive a set of equations for the velocity potential which are valid in channels of arbitrary cross section shape. At low Reynolds numbers, and when the channel is very long compared to its characteristic transverse dimension, the Navier-Stokes equation reduces to a single partial differential equation for a velocity potential in two space dimensions; the inhomogenous biharmonic equation with constant forcing.  This equation, which is derived in Sec.~\ref{sec:2}, has been widely studied in the literature as is describes the transverse displacement of a simply suspended thin elastic plate under uniform load. In Sec.~\ref{sec:3} we provide analytic solutions to four cases of porous channel flows in geometries where the solution of the corresponding elastic problem is known: Flow in a cylindrical tube, between parallel plates, in a triangular channel, and in a rectangular channel.

\section{Flow in channels with porous walls}
\label{sec:flowsketch}
\begin{figure}
  \centerline{\includegraphics[width=\textwidth]{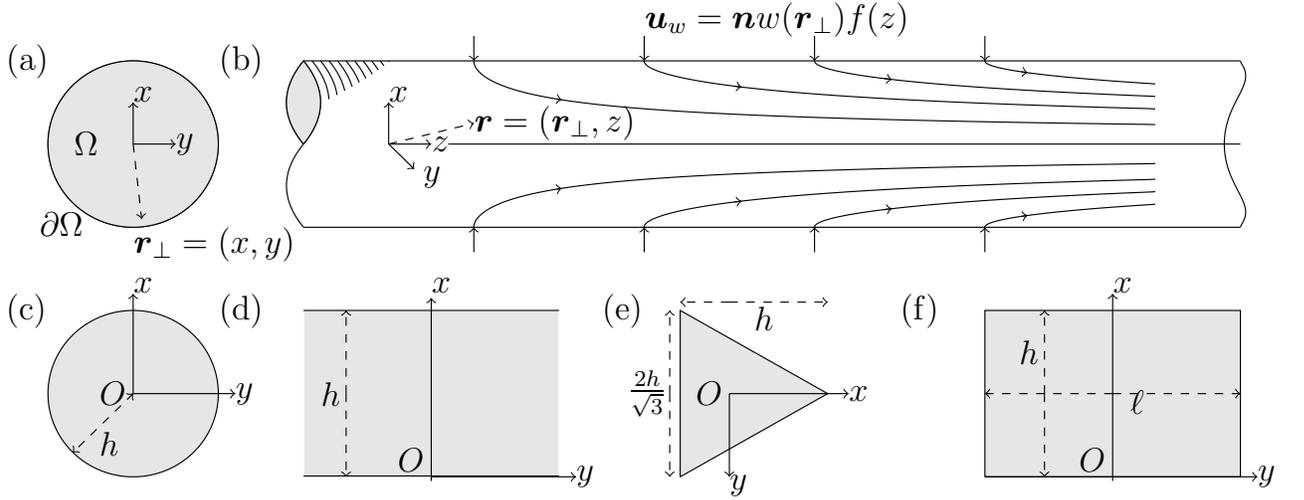}}
  \caption{Flow in channels with porous walls. (a) Schematic view of the channel cross section geometry. (b) Schematic view of the axial flow geometry. See Sec.~\ref{sec:flowsketch} for further details. (c-f) Sketch of geometries considered in Sec.~\ref{sec:FlowSolutions}. In (a) and (c-f) the $z$-axis is pointing into the page.}
\label{fig:flow_sketch}
\end{figure}
\label{sec:2}
We consider a long, straight channel parallel to the $z$-direction, and assume that it is translational invariant along this axis with an arbitrary, but constant, cross section $\Omega$ as shown in Fig.~\ref{fig:flow_sketch}(a-b). The channel has length $L$, perimeter $S$, cross section area $A$, and volume $V = LA$. The coordinates in the transverse $xy$-plane are denoted $\rperpbf=(x,y)$, so that the full coordinates are written as $\boldsymbol r=\rperpbf+\ez z$ and likewise for the gradient operator $\nablabf$, Laplace operator $\nabla^2$, and velocity field $\ubf$
\begin{align*}
\rbf 		&= \rperpbf+\ez z				&\text{where}	\qquad\rperpbf 			&= (x,y)\\
\nablabf 	&= \nablabf_\perp+\ebf_z\partial_z 	&\text{}	\qquad\nablabf_\perp 		&=(\partial_x,\partial_y) \\
\nabla^2 	&= \nabla^2_\perp+\partial^2_z		&\text{}	\qquad \nablabf^2_\perp 	&=\partial^2_x +\partial^2_y\\
\ubf		&= \ubf_\perp+\ez u_z			&\text{}	\qquad\ubf_\perp 			&= (u_x,u_y)
\end{align*}
where we have used the short hand notation $\partial_x f= \partial f / (\partial x)$. 

We consider the case of incompressible Newtonian fluids of viscosity $\eta$ and density $\rho$ (in the laminar regime) which are governed by the Navier-Stokes equation
\beqa
\rho\left(\partial_t \ubf +(\ubf\boldsymbol \cdot\nablabf)\ubf\right) &=& -\nabla p+\eta \nabla^2 \ubf\label{eq:NavierStokes}\label{eq:21}\\
\nablabf \boldsymbol \cdot \ubf &=& 0\label{eq:22}
\eeqa
where $p$ is the pressure. Writing the velocity field $\ubf$ in decomposed form as $\ubf=\ubf_\perp + \ez u_z$ Eqns.~\eqref{eq:21} and \eqref{eq:22} are
\beqa
\rho\left(\partial_t u_z +(\ubf_\perp\boldsymbol \cdot\nablabf_\perp+ u_z\partial_z)u_z\right) &=& -\partial_z p+\eta (\nabla^2_\perp+\partial^2_z) u_z\label{eq:23}\\
\rho\left(\partial_t \ubf_\perp +(\ubf_\perp\boldsymbol \cdot\nablabf_\perp+ u_z\partial_z) \ubf_\perp\right) &=& -\nablabf_\perp p+\eta (\nabla^2_\perp+\partial^2_z) \ubf_\perp\label{eq:24}\\
\nablabf_\perp\boldsymbol \cdot \ubf_\perp+\partial_z u_z &=& 0\label{eq:25}
\eeqa
We assume that the flow is driven by a prescribed injection or suction of fluid through the porous wall which leads to a normal flow velocity component at the channel wall $u_w(\rbf)= w(\rbf_\perp)f(z)$ of typical magnitude $u_0$. The boundary conditions thus require that the tangential velocity component vanishes on the channel wall and that the normal velocity component is $u_w(\rbf)$
\begin{align}
u_z(\rbf)&=0, &\rbf \in \partial \Omega\label{eq:bc01},\\
\tbf\boldsymbol \cdot\ubf_\perp(\rbf) &= 0, &\rbf \in \partial \Omega\label{eq:bc03},\\
\nbf\boldsymbol \cdot\ubf_\perp (\rbf) &=u_w(\rbf)=w(\rbf_\perp)f(z),&\rbf \in \partial \Omega,\label{eq:bc02}
\end{align}
where $\nbf$ is a outward pointing normal unit vector and $\boldsymbol t$ is a tangent unit vector to the boundary $\partial\Omega$ in the $\rbf_\perp$ plane.
\subsection{Non-dimensionalization}
To cast Eqns.~\eqref{eq:23}-\eqref{eq:25} into a simpler form we non-dimensionalize the equations using the characteristic wall flow velocity $u_0$, channel length $L$, and transverse dimension $a= A/S$:
\begin{align*}
z&= L z'  & 
\rbf_\perp&=a\rbf_\perp' 
&t&=\frac{a}{u_0}t'
&\ubf_\perp &= u_0 \ubf_\perp' & 
u_z &= \frac{ L}{a}u_0u_z' & 
p&=\frac {L^2\eta u_0}{a^3}p'
\\
\nablabf_\perp&=\frac{1}{a}\nablabf_\perp'&
 \nabla_\perp^2 &=\frac{1}{a^2}\nabla'_{\perp}{}^{2} & 
 \partial_z&=\frac 1L\partial_{z'}
&\ppartial_z &=\frac{1}{L^2}\ppartial_{z'}&
\partial_t &=\frac{u_0}{a}\partial_{t'} &
\end{align*}
With this change of variables, Eqns.~\eqref{eq:23}-\eqref{eq:25} are
\beqa
\frac{\rho u_0a}{\eta}\left(\partial_{t'} u_z' +(\ubf_\perp'\boldsymbol \cdot\nablabf_\perp'+ u_z'\partial_{z'})u_z'\right) &=& -\partial_{z'} p'+ \left(\nabla^2_\perp{}'+\left(\frac{a}{L}\right)^2\partial^2_{z'}\right) u_z'\\
\frac{\rho u_0a^3}{\eta L^2}\left(\partial_t \ubf_\perp' +(\ubf_\perp'\boldsymbol \cdot\nablabf_\perp'+ u_z'\partial_{z'}) \ubf_\perp'\right) &=& -\nablabf_\perp' p'+\left(\frac{a}{L}\right)^2\left(\nabla^2_\perp{}'+\left(\frac{a}{L}\right)^2\partial^2_{z'}\right) \ubf_\perp'\nonumber\\\\
\nablabf_\perp'\boldsymbol \cdot \ubf_\perp'+\partial_{z'} u_z' &=& 0
\eeqa
Introducing the Reynolds number $\Rey=\frac{\rho u_0a}{\eta}$ based on the wall velocity $u_0$ and the aspect ratio $\alpha=\frac{a}{L}$, and dropping the primes for ease of reading we finally have that
\beqa
\Rey\left(\partial_{t} u_z +(\ubf_\perp\boldsymbol \cdot\nablabf_\perp+ u_z\partial_{z})u_z\right) &=& -\partial_{z} p+ \left(\nabla^2_\perp+\alpha^2\partial^2_{z}\right) u_z\label{eq:NavierStokesNondim00}\\
\alpha^2\Rey\left(\partial_t \ubf_\perp +(\ubf_\perp\boldsymbol \cdot\nablabf_\perp+ u_z\partial_{z}) \ubf_\perp\right) &=& -\nablabf_\perp p+\alpha^2\left(\nabla^2_\perp+\alpha^2\partial^2_{z}\right) \ubf_\perp\label{eq:NavierStokesNondim01}\\
\nablabf_\perp\boldsymbol \cdot \ubf_\perp+\partial_{z} u_z &=& 0\label{eq:NavierStokesNondim02}
\eeqa
with the boundary conditions given in Eqns.~\eqref{eq:bc01}-\eqref{eq:bc02}.
\subsection{Similarity solutions}
The form of the boundary conditions is such that the in-plane velocity $\ubf_\perp$ could be irrotational and proportional to $f(z)$ everywhere (see Eqns.~\eqref{eq:bc03}-\eqref{eq:bc02}), while the axial velocity $u_z$ should be proportional to the total volume of liquid which has entered the channel at $z'<z$, \ie $\ubf_\perp \propto F(z) =\int_{0}^z f(z') \ud z' + F_0$. It is worthwhile to enquire if the differential equation permits solutions of this form, and we therefore write the velocity $\ubf$ as a similarity solution
\begin{eqnarray}
\ubf_\perp&=&-f(z)\boldsymbol g(\rbf_\perp)\label{eq:uperp1}\label{eq:potential00}\\
u_z&=&F(z)h(\rbf_\perp)\label{eq:uz1}\
\end{eqnarray}
where $\boldsymbol g(\rbf_\perp)$ and $h(\rbf_\perp)$ are unknown functions of the radial position $\rbf_\perp$ only. Using Eq.~\eqref{eq:potential00}, we find from the continuity equation \eqref{eq:NavierStokesNondim02}
\beq
\nabla_\perp \boldsymbol \cdot \boldsymbol g = h.
\label{eq:potential01}
\eeq
Since $\ubf_\perp$, and therefore $\boldsymbol g$, is irrotational by assumption, we may write $\boldsymbol g=\nablabf_\perp \phi$ where $\phi$ is a velocity potential. Eq.~\eqref{eq:potential01} then implies that $h=\nabla^2 \phi$. The velocity field we seek is thus of the form
\begin{eqnarray}
\ubf_\perp&=&-f(z)\nablabf_\perp \phi(\rbf_\perp)\label{eq:uperp1}\\
u_z&=&F(z)\nabla_\perp^2 \phi(\rbf_\perp)\label{eq:uz1}\
\end{eqnarray}
Substituting Eqns.~\eqref{eq:uperp1} and \eqref{eq:uz1} into Eqns.~\eqref{eq:NavierStokesNondim00} and \eqref{eq:NavierStokesNondim01} we find 
\beqa
\Rey \left[\partial_t\left(F\nabla_\perp\phi\right)+fF\left((\nabla_\perp^2\phi)^2-(\nablabf_\perp\phi\boldsymbol\cdot\nablabf_\perp)\nabla_\perp^2\phi\right)\right]&=&-\partial_z p+F\nabla_\perp ^4+\alpha^2\partial_z f\nabla_\perp^2\phi\nonumber \\\label{eq:220}\\
\alpha^2\Rey\left[-\partial_t\left(f\nablabf_\perp \phi\right)+f^2(\nablabf_\perp\phi\boldsymbol\cdot\nablabf_\perp)\nablabf_\perp\phi-F\partial_z f\nabla^2_\perp\phi \nablabf_\perp\phi\right]
&=&
-\nablabf_\perp p-\alpha^2\left(f\nabla_\perp^2 \nablabf_\perp \phi-\alpha^2\partial^2_z f\nablabf_\perp\phi \right)
\nonumber \\\label{eq:221}
\eeqa
The boundary conditions for $(\rbf_\perp,z) \in \partial \Omega$ in Eqns.~\eqref{eq:bc01}-\eqref{eq:bc03} are
\beqa
F(z)\nabla^2_\perp \phi(\rbf_\perp)&=&0\label{eq:bc001}\\
-(\tbf\boldsymbol \cdot\nablabf_\perp\phi(\rbf_\perp,z))f(z)&=&0,\label{eq:bc003}\\
-(\nbf\boldsymbol \cdot\nablabf_\perp\phi(\rbf_\perp))f(z)&=&w(\rbf_\perp)f(z)\label{eq:bc002}
\eeqa
These boundary condition may be considerably simplified by noting that Eq.~\eqref{eq:bc003} implies that $\phi(\rbf_\perp,z)= k$ is constant on the boundary $\partial \Omega$. (If $\partial \Omega$ consists of several physically separate boundaries, $k$ may take on different values on each of these). With this, Eqns.~\eqref{eq:bc001}--\eqref{eq:bc002} become
\begin{align}
\ppartial_n \phi&=0\label{eq:bc001a}\\
\phi&=k\label{eq:bc003c}\\
-\partial_n\phi &=w\label{eq:bc002bb}
\end{align}
where we have assumed that $f(z)\neq 0$, $F(z)\neq 0$, and use the notations $\partial_n$ and $\partial^2_n$ for first and second order normal derivatives.
\subsection{The case $\Rey\ll 1$ and $\alpha\ll 1$ }
\label{sec:3}
When both the Reynolds number $\Rey\ll 1$ and aspect ratio $\alpha\ll1 $ are small, we find from Eqns.~\eqref{eq:220}--\eqref{eq:221} that
\beqa
{\partial_z p}{}&=&F\nabla_\perp^4\phi,\label{eq:lowreypsi}\\
\nablabf_\perp p &=& 0.\label{eq:lowreypsi02}
\eeqa
Eq.~\eqref{eq:lowreypsi02} implies that the pressure is a function of $z$ only, and we may write $p=p(z)$.
By introducing $P(z)={\partial_z p_z}/{F}$, we find the governing equation for the velocity potential $\phi$
\beq
\nabla_\perp^4\phi=P\label{eq:001}\\
\eeq
with the boundary conditions given in Eqns.~\eqref{eq:bc001a}-\eqref{eq:bc002bb}.
\subsection{Analogy with the theory of simply suspended plates}
The equation of motion for a thin suspended plate under a uniform transverse load $q$ is the inhomogeneous biharmonic equation
\beq
\nablabf_\perp^4 W  = \frac{q}{D},\label{eq:661}
\eeq
where $W$ is the displacement at any point from the position of equilibrium and $D$ depends only on the tension and mass of the plate.
This equation is of the same form as Eq.~\eqref{eq:001}, and the amplitude of the displacement may be taken to represent $\phi$ if
\beq
P=\frac{q}{D}.
\eeq
It appears therefore that if a solution of the problem of a suspended plate has been obtained, a problem of viscous motion in a porous tube has also been solved. This analogy is exact, so long as the boundary conditions in Eq.~\eqref{eq:bc001a}--\eqref{eq:bc002bb} are also fulfilled in the plate problem. This occurs when the plate is simply suspended, i.e. when on the boundary $\partial^2W/\partial n^2=0$ and $W=k$, in which case Eqns.~\eqref{eq:bc001a} and \eqref{eq:bc003c} are fulfilled. The final boundary condition (Eq.~\eqref{eq:bc002bb}) which determines the angle of deflection $\partial W/\partial n$ or the normal flow velocity across the membrane $u_w$ is set by the choice of $P=q/D$ if $\Omega$ is simply connected or by $P$ and $k$ on each of the boundaries if $\Omega$ is multiply connected.

This observation that Eqns.~\eqref{eq:661} and \eqref{eq:001} are of the same form adds to a list of viscous flow problems which may be solved by studying displacements and vibrations of thin plates (see e.g. \citet{Rayleigh1893a,Taylor1923,Richards1960,Society2012,Lauga2004}).
%
\section{Flow solutions}
\label{sec:FlowSolutions}
Solutions of the inhomogeneous biharmonic equation \eqref{eq:001} must satisfy the governing differential equation and boundary conditions characterizing each geometry. The fulfillment of the boundary conditions often presents considerable mathematical difficulties and thus in general, analytic solutions are rare. 
Taking advantage of the solutions known from plate bending theory (see e.g. \citet{Timoshenko1964,Ventsel2001}), we are able to provide analytic solutions to porous channel flows in geometries where the solution of the corresponding elastic problem is known, such as in channels of rectangular and triangular cross section shape. In these cases the solution relies on either a parametrization of the boundary or a series solution.  Before we consider these three dimensional cases, however, we illustrate the solution technique on a few two dimensional flow problems which have been solved by other means in the literature.
\subsection{Flow in a cylindrical tube}
\begin{figure}
  \centerline{\includegraphics[width=6cm]{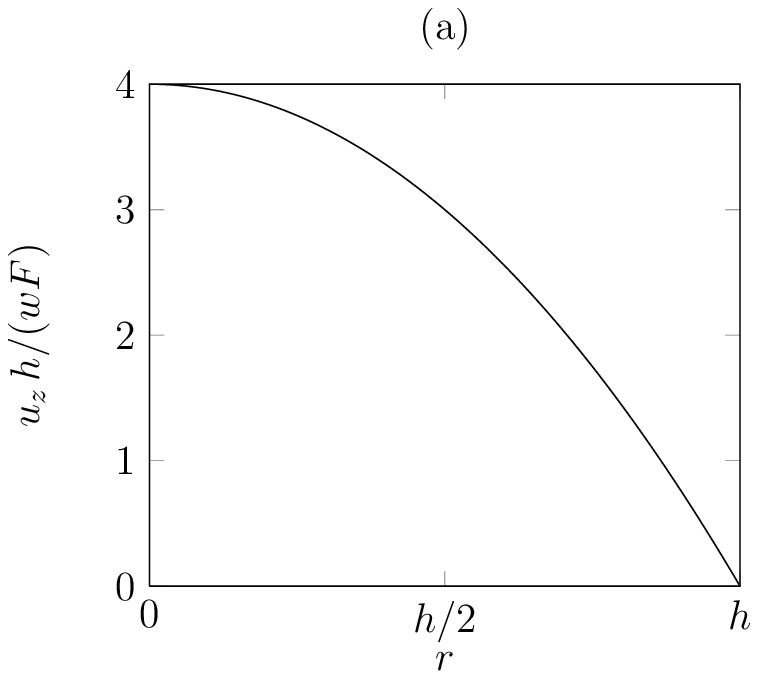}\hfill
  \includegraphics[width=6.4cm]{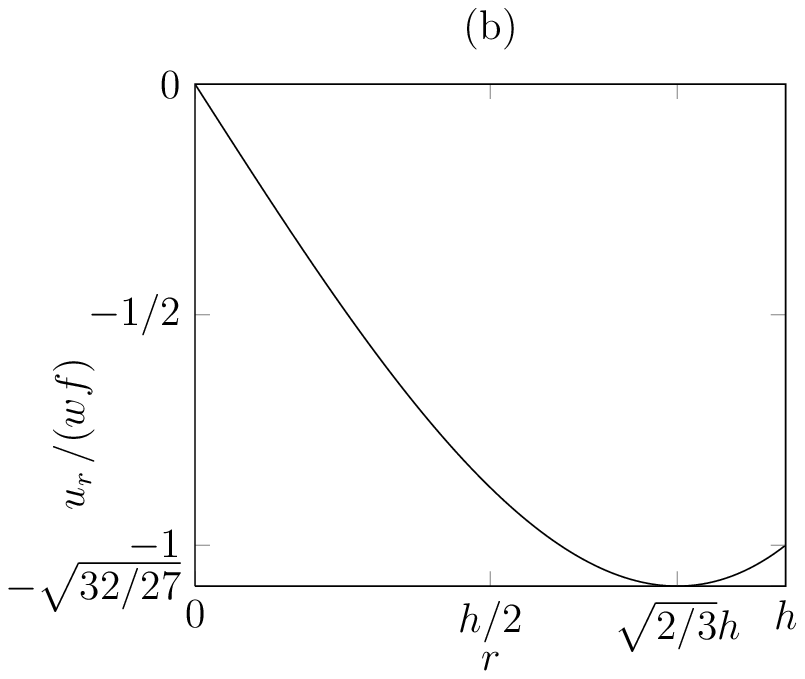}}
  \caption{Flow in a cylindrical tube of radius $h$. (a) Axial flow velocity $u_z$ given in Eq.~\eqref{eq:cyl02} plotted as a function of radial position $r$. (b) Radial flow velocity $u_r$ given in Eq.~\eqref{eq:cyl01} plotted as a function of radial position $r$. Note that the radial velocity has a minimum at $r=\sqrt{2/3}h\simeq 0.817\, h$ where the value is $u_r(\sqrt{2/3}h)=-\sqrt{{32}/{27}}wf\simeq-1.09\, wf(z)$.}
\label{fig:cylinder}
\end{figure}
Consider the flow in a cylindrical tube of radius $h$ as sketched in Fig.~\ref{fig:flow_sketch}(c). Let the normal velocity at the wall be $u_r(h)=-wf(z)$. Assuming rotational symmetry of the velocity field, the governing equation \eqref{eq:001} is
\beq
\left(\partial_r^4 +\frac{2}{r}\partial_r^3+\frac{1}{r^2}\partial_r^2+\frac{1}{r^3}\partial_r\right)\phi = P
\label{eq:phicylinder}
\eeq
which has the solution
\beq
\phi = \frac{P}{64} r^4+\frac 14 (2B_2-B_3)r^2+(B_1+\frac 12 r^2 B_3)\log r + B_4.
\eeq
With $k=0$ at the $r=h$ channel wall, the boundary conditions in Eqns.~\eqref{eq:bc001a}-\eqref{eq:bc002bb} are
\beq
\frac 1r\partial_r(r\partial_r \phi(h))=0,\quad \phi(h)=0,\quad
-\partial_r\phi(h)=-w
\eeq
Assuming further that the radial flow component vanishes at $r=0$ ($u_r(0)=0$) such that $-\partial_r\phi(0)=0$ we find for the for the velocity potential $\phi$ in Eq.~\eqref{eq:phicylinder}
\beq
\phi(r)=\left(-\frac{3 h}{4}+\frac{r^2 }{h}-\frac{r^4}{4 h^3}\right)w.
\eeq
The radial and axial velocity components can now be found from Eqns.~\eqref{eq:uperp1}-\eqref{eq:uz1}
\beqa
u_r&=&-f\partial_r \phi=\frac{r \left(r^2-2 h^2\right) w}{h^3}f\label{eq:cyl01},\\
u_z&=&F\frac 1r\partial_r(r\partial_r \phi)=\frac{4 (h^2-r^2)w}{h^3}F\label{eq:cyl02}
\eeqa
in agreement with \eg \citet{Aldis1988a}. The two velocity components are plotted in Fig.~\ref{fig:cylinder}. We note that the radial velocity has a minimum at $r=\sqrt{2/3}h\simeq 0.817\, h$ where the value is $u_r(\sqrt{2/3}h)=-\sqrt{{32}/{27}}wf\simeq-1.09\, wf(z)$.

\subsection{Flow between parallel plates}
\label{sec:plates}
\begin{figure}
  \begin{center}
  \includegraphics[width=6cm]{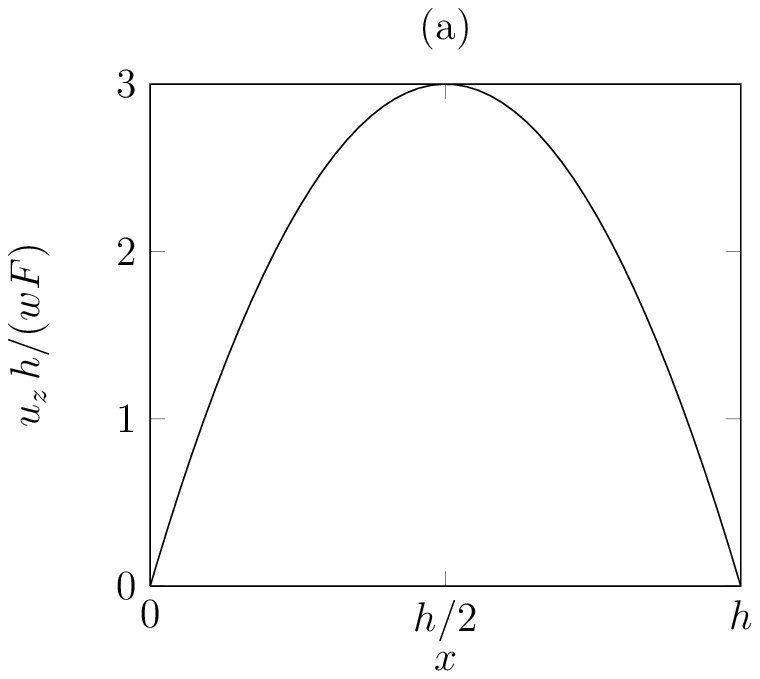}
  \hfill\includegraphics[width=6cm]{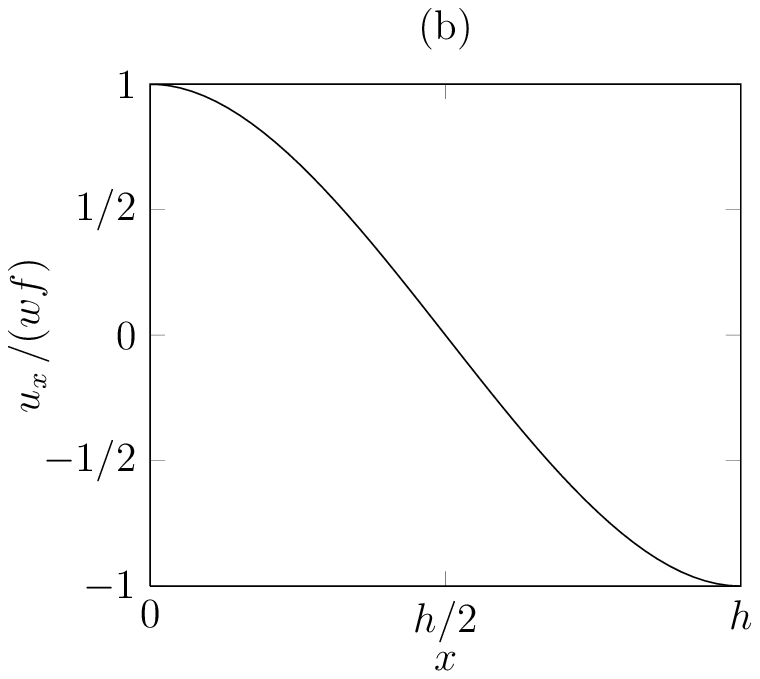}
  \caption{Flow between parallel porous plates located at $x=0$ and $x=h$ with boundary conditions $u_x(0)=wf$ and $u_x(h)=-wf$.  (a) Axial flow velocity $u_z$ given in Eq.~\eqref{eq:pp02} plotted as a function of vertical position $x$. (b) Transverse flow velocity $u_x$ Eq.~\eqref{eq:pp01} plotted as a function of vertical position $x$.
\label{fig:parallel_plates}}
\end{center}
\end{figure}
Consider the flow between two infinite parallel plates located at $x=0$ and $x=h$ as sketched in Fig.~\ref{fig:flow_sketch}(d). Let the normal velocity at the walls be $u_x(0)=w_-f(z)$ and  $u_x(h)=w_+f(z)$. The governing equation \eqref{eq:001} is
\beq
\partial_x^4\phi = P\\
\eeq
which has the  solution
\beq
\phi =A_0+A_1x+A_2x^2+A_3x^3 +  \frac{P}{24}x^4.
\eeq
Applying the boundary conditions in Eqns.~\eqref{eq:bc001a}-\eqref{eq:bc002bb}
\beq
\partial_x^2\phi(0)=0,\quad
\partial_x^2\phi(h)=0,\quad
\phi(0)=0,\quad
-\partial_x\phi(0)=w_-,\quad\text{and}\quad
-\partial_x\phi(h)=w_+,\quad 
\eeq
yields
\beq
\phi =\frac{(w_--w_+) x^3}{h^2}+\frac{(w_+-w_-) x^4}{2 h^3}-w_- x.
\eeq
The transverse and axial velocity components can be found from Eqns.~\eqref{eq:uperp1}-\eqref{eq:uz1}
\beqa
u_x&=&-f(z)\partial_x\phi=\left(\frac{3(w_+-w_-) x^2}{h^2}+\frac{2(w_--w_+) x^3}{h^3}+w_-\right)f\label{eq:pp01}\\
u_z&=&F(z)\partial_x^2\phi=\frac{6(w_--w_+)(h-x)x}{h^3}F\label{eq:pp02}
\eeqa
The solution for the special case $w_ -=-w_+=w$ is shown in 
Fig.~\ref{fig:parallel_plates}. In that case the velocity components are
\beqa
u_x&=&\frac{h^3-6hx^2+4x^3}{h^3}wf,\label{eq:pp01}\\
u_z&=&\frac{12 (h-x)x}{h^3}wF.\label{eq:pp02}
\eeqa
These were first by obtained \citet{Berman1953} who considered the case $f=1$.
\subsection{Flow in an equilateral triangle}
%
\begin{figure}
\begin{center}
\hspace{-0.6cm}\includegraphics[width=13cm]{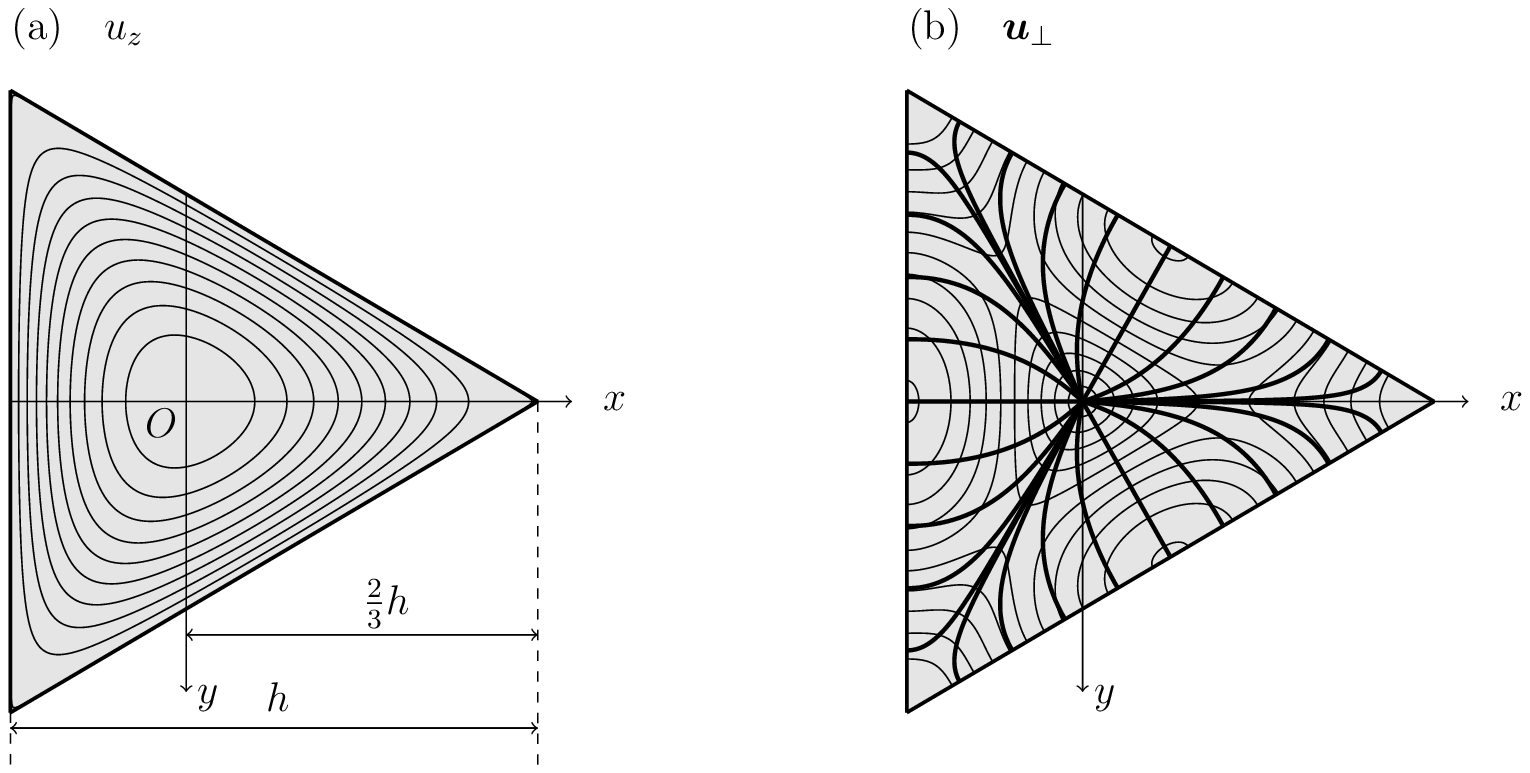}\hfill
\includegraphics[width=6cm]{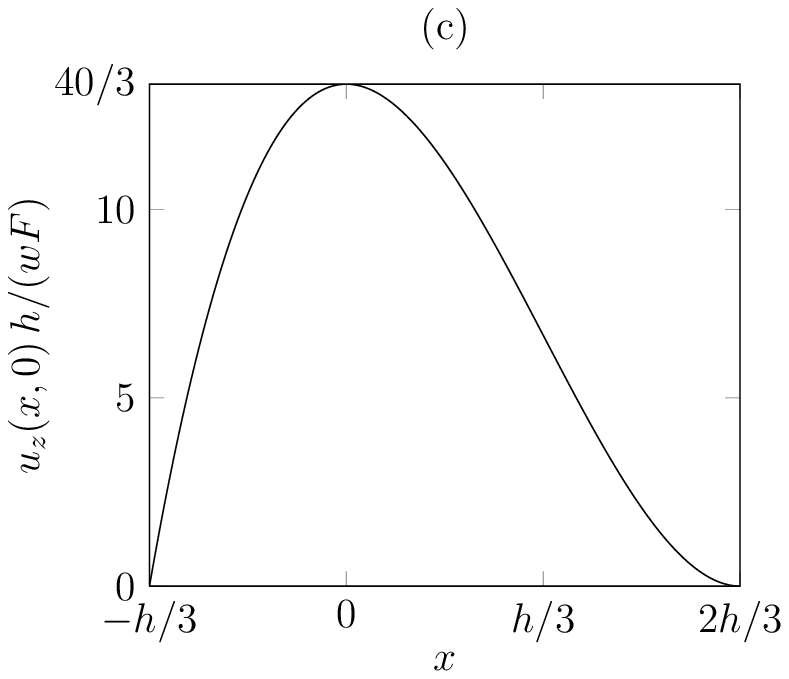}\hfill
\includegraphics[width=6cm]{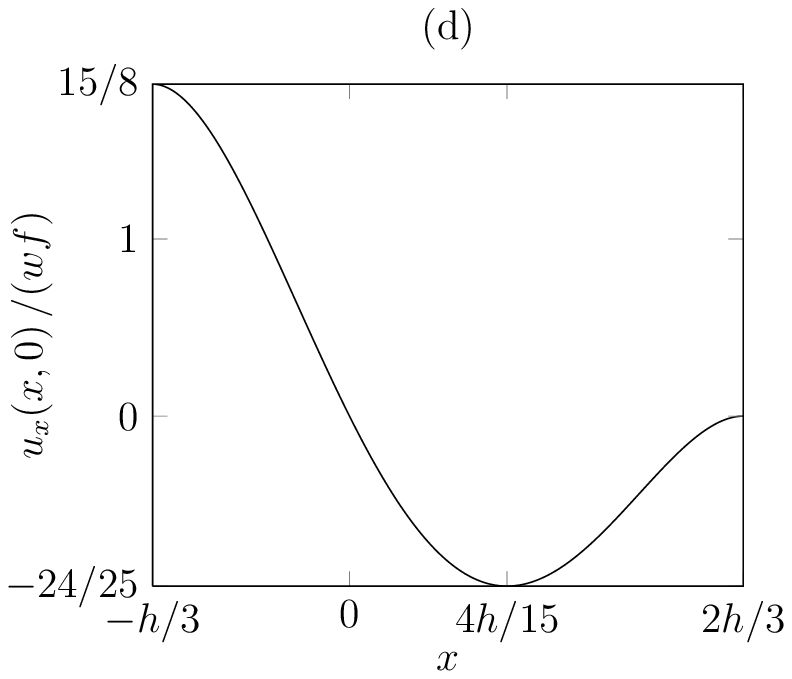}\hfill
\caption[]{
Flow in a porous tube whose cross-section is an equilateral triangle of height $h$ and side length $2h/\sqrt 3$. (a) Contour lines for the axial flow velocity field $u_z(x,y)$. The contour lines are shown in steps of $10\%$ of the maximum value. (b) Stream(thick)- and contour(thin) lines for the transverse velocity field $\ubf_\perp=(u_x,u_y)$. The contour lines show the velocity magnitude $|\ubf_\perp|$ in steps of $10\%$ of the maximum value. (c) A plot of $u_z(x,0)$ along the centerline parallel to the $x$-axis. (d) A plot of $u_x(x,0)$ along the centerline parallel to the $x$-axis.\label{fig:triangle}}
\end{center}
\end{figure}
%
Consider the flow of liquid in a porous channel whose cross-section is an equilateral triangle of height $h$ and side length $2h/\sqrt 3$, as shown in Fig.~\ref{fig:flow_sketch}(e). Let the normal wall velocity be given by $u_w=w(x,y)f(z)$, and let the mean value of $w(x,y)$ when averaged over the channel wall be $w_0$. The governing equation \eqref{eq:001} is
\beq
(\ppartial_x +\ppartial_y)^2 \phi = P,
\label{eq:biharmcart}
\eeq
while the boundary conditions in Eqns.~\eqref{eq:bc001a}-\eqref{eq:bc002bb} are
\begin{equation}
\ppartial_n \phi = 0,
\quad \phi = 0, \quad\text{and}\quad
\quad-\partial_n\phi = w.\label{eq:bc003b}
\end{equation}
The solution of Eq.~\eqref{eq:biharmcart} with the boundary conditions $\ppartial_n \phi = 0$ and $\phi = 0$
is
\beq
\phi = \frac{P}{64h}\left(x^3-3y^2x-h(x^2+y^2)+\frac{4}{27}h^3\right)\left(\frac 49h^2-x^2-y^2\right)
\label{eq:phitria}
\eeq 
The magnitude of the wall normal flow velocity $u_w=w(x,y)f(z)$ is the same on each of the three boundaries. At the $x=-{h}/{3}$ wall we find
\beq
u_x(-h/3,y)=\left.-f\partial_x \phi(x,y)\right|_{x=-h/3}=-\frac{P \left(h^2-3 y^2\right)^2}{192 h}f.\label{eq:trianglewallvelocity}
\eeq
The mean velocity (when averaged over the $x=-h/3$ boundary) is $-{h^3 Pf}/{360}$
so by choosing $P=-{360w_0}/{(h^3)}$, the average inflow velocity becomes $w_0f$.
Using $P=-{360w_0}/{(h^3)}$ in Eq.~\eqref{eq:phitria}, the velocity components can be found from  Eqns.~\eqref{eq:uperp1}-\eqref{eq:uz1}
recalling that $u_x = -f\partial_x\phi$, $u_y = -f\partial_y\phi$ and $u_z=F(\ppartial_x + \ppartial_y)\phi$
\beqa
u_x&=&\frac{5 \left(18 \left(6 h x+9 x^2-2 h^2\right) y^2+81 y^4- x (8 h+15 x)(2 h-3 x)^2\right) }{24 h^4}w_0f\\
u_y&=&-\frac{5 (h+3 x) y \left(8 h^2-6 a x-9 \left(x^2+3 y^2\right)\right)}{6 h^4}w_0f\\
u_z&=&\frac{10 (h+3 x) \left((2 h-3 x)^2-27 y^2\right)}{3 h^4}w_0F
\eeqa
The velocity field is shown in Fig.~\ref{fig:triangle}.
\subsection{Flow in an rectangular channel}
%
\begin{figure}
\begin{center}
\hspace{-0.61cm}\includegraphics[width=14cm]{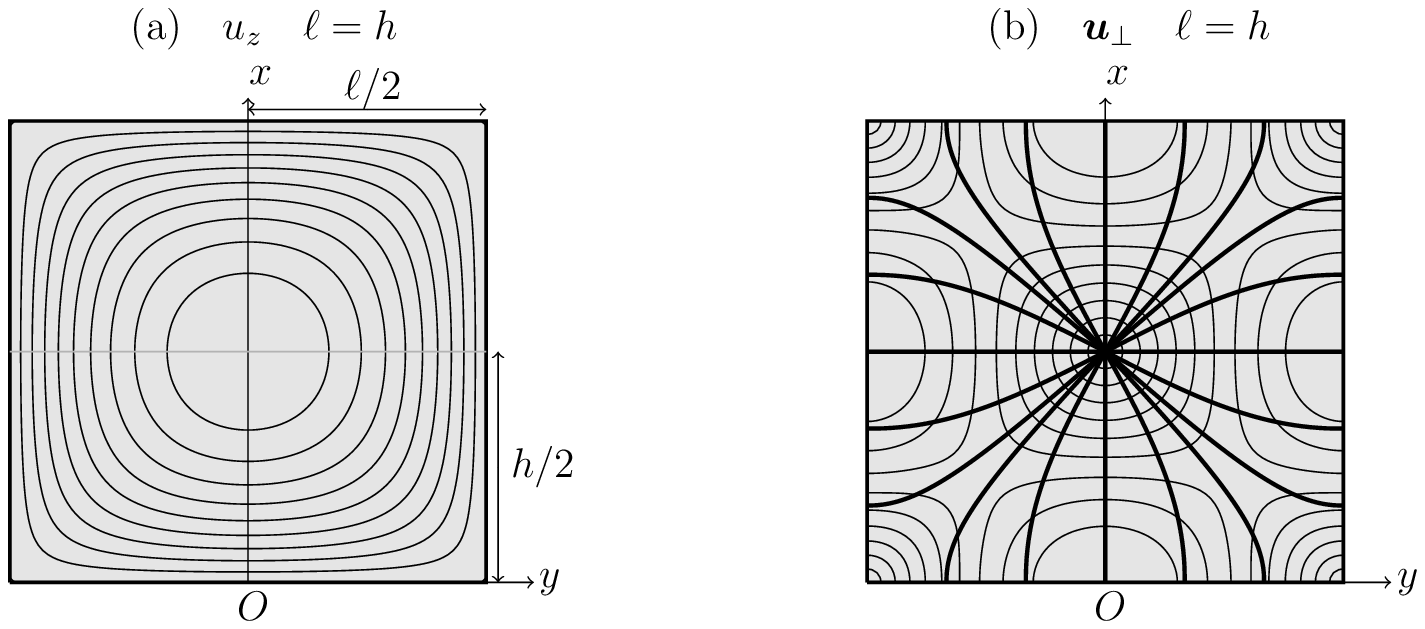}\vspace{-1cm}\hfill
\includegraphics[width=6cm]{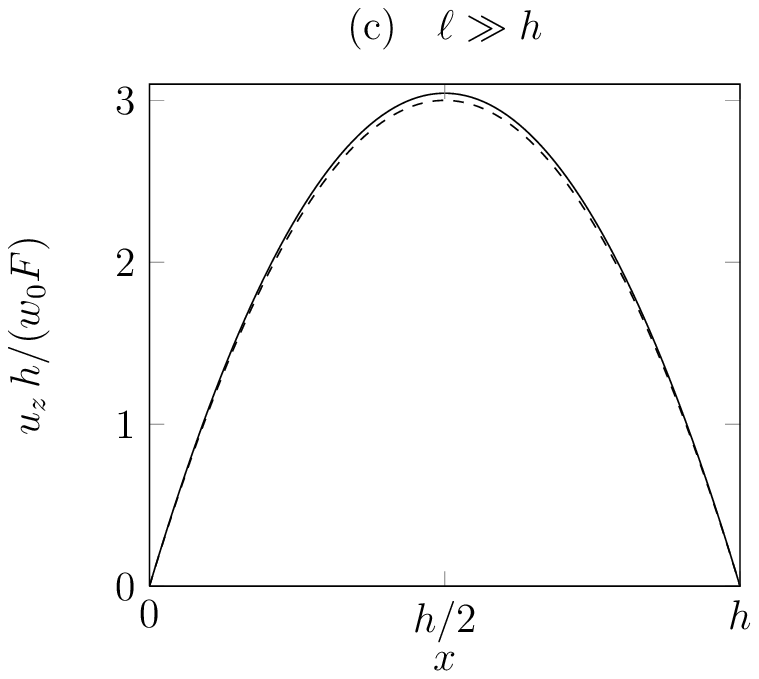}\hfill
\includegraphics[width=6cm]{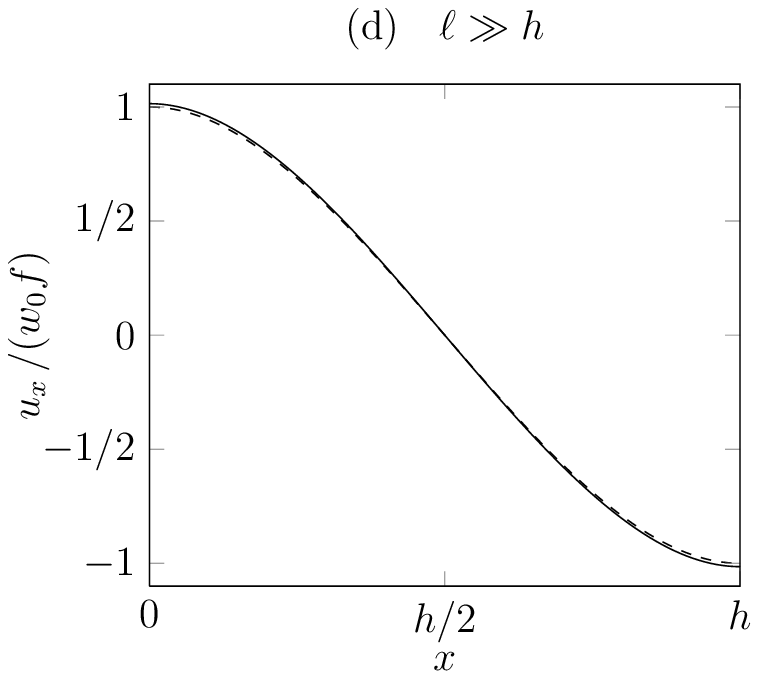}\hfill
\caption[]{Flow in a rectangular porous tube with sides of length $h$ and $\ell$. (a) Contour lines for the axial flow velocity field $u_z(x,y)$ for the case $h=\ell$. The contour lines are shown in steps of $10\%$ of the maximum value. (b) Stream(thick)- and contour(thin) lines for the transverse velocity field $\ubf_\perp$ for $h=\ell$. The contour lines show  the velocity magnitude $|\ubf_\perp|$ in steps of $10\%$ of the maximum value. (c) A plot of $u_z(x,0)$ (solid line) along the centerline parallel to the $x$-axis for $\ell/h=1000$. Dashed line shows $u_z(x,0)$ for parallel plates, c.f. Eq.~\eqref{eq:pp02}. (d) A plot of $u_x(x,0)$ (solid line) along the centerline parallel to the $x$-axis $\ell/h=1000$. Dashed line shows $u_x(x,0)$ for parallel plates, c.f. Eq.~\eqref{eq:pp01}. In (a) and (b), the first $100$ terms in Eqns.~\eqref{eq:rectux}--\eqref{eq:rectuz} were used.\label{fig:rectangle}}
\end{center}
\end{figure}
%
Consider the flow in a porous channel whose cross-section is a rectangle with sides of length $h$ and $\ell$ centered at $(x,y)=(0,h/2)$ as shown in Fig.~\ref{fig:flow_sketch}(d). Let the normal wall velocity be given by $u_w=w(x,y)f(z)$, and let the mean value of $w(x,y)$ when averaged over the channel walls be $w_0$.

The equations of motion and boundary conditions are once again given by Eqns.~\eqref{eq:biharmcart}-\eqref{eq:bc003b}. As we need to evaluate the first and second order $x$ and $y$-derivatives to determine the velocity field, Levy's solution of  the corresponding plate bending problem is convenient
\beqa
\phi 
\label{eq:levy}
&=& \frac{4Ph^4}{\pi^5}\sum_{n=1,3,\ldots} \frac{1}{n^5}\left(1-\mathcal A_n\cosh\frac{2\alpha_n y}{\ell}+\mathcal B_n y\sinh \frac{2\alpha_ny}{\ell}\right)\sin \frac{n\pi x}{h}
\label{eq:levy}
\eeqa
where $\alpha_n ={n\pi \ell}/({2h})$,  $\mathcal A_n=({\alpha_n\tanh \alpha_n +2})/({2\cosh \alpha_n})$, and $\mathcal B_n={\alpha_n}/({\ell\cosh \alpha_n})$. 
From Eqns.~\eqref{eq:levy}, \eqref{eq:uperp1}, and \eqref{eq:uz1} we may calculate the velocity components 
\beqa
u_x&=&-\frac{4Ph^3}{\pi^4}f\sum_{n=1,3,\ldots} \frac{1}{n^4}\left(1-\mathcal A_n\cosh\frac{2\alpha_n y}{\ell}+\mathcal B_n y\sinh \frac{2\alpha_ny}{\ell}\right)\cos \frac{n\pi x}{h}\label{eq:rectux}\\
u_y&=&-\frac{4Ph^4}{\pi^5}f\sum_{n=1,3,\ldots}\frac{1}{n^5}\left(-\frac{2\alpha_n}{\ell}\mathcal A_n\sinh \frac{2\alpha_n y}{\ell}+\mathcal B_n\left(\sinh \frac{2\alpha_n y}{\ell}+\frac{2\alpha_n}{\ell} y\cosh\frac{2\alpha_n y}{\ell}\right)\right)\sin \frac{n\pi x}{h},\nonumber\\\label{eq:rectuy}\\
u_z&=&-\frac{4Ph^2}{\pi^3}F\sum_{n=1,3,\ldots} \frac{1}{n^3}\left(1-\frac{\cosh \frac{n\pi y}{h}}{\cosh\alpha_n}\right)\sin \frac{n\pi x}{h}\label{eq:rectuz}
\eeqa
To determine $P$ such what the average inflow velocity is $w_0f$, we solve for $P$ in
\beq
\frac{2}{h} \int_0^hu_y(-\ell/2,x,z)\ud x +\frac{2}{\ell} \int_{-\ell/2}^{\ell/2} u_x(0,y,z)\ud y = w_0f.\label{eq:findPrect}
\eeq
In the following we determine $P$ for the special cases $\ell = h$ and $\ell \gg h$.
\subsubsection{The case $\ell=h$}
To determine the velocity field in the case $\ell = h$ we note that the sums in Eqns.~\eqref{eq:rectux}--\eqref{eq:rectuz} converge very rapidly, and that the flow profile $u_x(0,y)$ at the $x=0$ wall is well approximately by the first term in Eq.~\eqref{eq:rectux}
\begin{eqnarray}
u_x(0,y,z)&\simeq&-\frac{4Ph^3}{\pi^4}\left(1-\mathcal A_n\cosh\frac{2\alpha_n y}{\ell}+\mathcal B_n y\sinh \frac{2\alpha_ny}{\ell}\right)f\label{eq:approx}
\end{eqnarray}
From Eq.~\eqref{eq:findPrect} we therefore find
\begin{equation}
\frac{4}{\ell}\int_{-\ell/2}^{\ell/2}u_x(0,y,z) \ud y\simeq -\frac{8h^3P}{\ell\pi^5}{\left(\sech\frac{\ell \pi }{2 h}\right)^2 \left(\ell \pi  \left(2+\cosh\frac{\ell \pi }{h}\right)-3 h \sinh\frac{\ell \pi }{h}\right)}f=w_0f\label{eq:rectuxmean},
\end{equation}
which determine $P$ as a function of the average normal flow velocity $w_0$
\beq
P=-\frac{\pi^5}{16}\frac{1+\cosh \pi}{\pi(2+\cosh \pi)-3\sinh \pi}\frac{w_0}{h^3}\simeq -29.90\frac{w_0}{h^3}\label{eq:rectPapprox}
\eeq
Summing the first $100$ terms in Eqns.~\eqref{eq:rectux}-\eqref{eq:rectuy} we find from Eq.~\eqref{eq:findPrect} that $P= -28.46\,{w_0}/{h^3}$ so the error in the expression for $P$ in Eq.~\eqref{eq:rectPapprox} is less than $5\%$. The velocity field for $\ell=h$ is shown in Fig.~\ref{fig:rectangle}(a-b).
\subsubsection{The case $\ell/h\gg 1$}
To determine the velocity field in the case $\ell \gg h$ we again use Eq.~\eqref{eq:approx}.
Near the centerline parallel to the $x$-axis, i.e. for $|y|\ll \ell$, we find from Eq.~\eqref{eq:approx} that the wall velocity $u(0,0,z)\simeq -{4Ph^3f}/{\pi^4}$ so with $P=-\pi^4w_0/(4h^3)$ the condition in Eq.~\eqref{eq:findPrect} is fulfilled.
%
For the transverse velocity $u_x$ along the centerline $y=0$ we obtain from Eq.~\eqref{eq:rectux}
\begin{equation}
u_x(x,0,z)=w_0f\sum_{n=1,3,\ldots} \frac{1}{n^4}\cos \frac{n\pi x}{h}.
\end{equation}
This series can be summed
\begin{equation}
u_x(x,0,z)=\frac{\pi^4}{96}\frac{h^3-6hx^2+4x^3}{h^3} w_0 f.\\
\end{equation}
The prefactor $\pi^4/96\simeq 1.01$, in good agreement with Eq.~\eqref{eq:pp01}. Similarly, we find from Eq.~\eqref{eq:rectuz} 
\begin{equation}
u_z(x,0,z)=\frac{\pi^4}{8}\frac{(h-x)x}{h^3}w_0F
\end{equation}
The prefactor ${\pi^4}/{8}\simeq 12.18$  in good agreement with Eq.~\eqref{eq:pp02}.
The velocity field for $\ell\gg h$ is shown in Fig.~\ref{fig:rectangle}(c-d).
\section{Conclusion}
We have analyzed slow flow in channels with porous walls. A similarity transformation reduces the Navier-Stokes equations to a set of coupled equations for the velocity potential in two dimensions.  We have shown that when the Reynolds number and channel aspect ratio is small, an analogy exists between flow in channels with porous walls and bending of simply suspended plates under uniform load. 
If a solution of the problem of a suspended plate has been obtained, a problem of viscous motion in a porous tube has thus also been solved. We have applied this result to flow in rectangular and triangular channels. Our results provide a general framework for the extension of Berman flow \citep{Berman1953} to three dimensions.
\section{Adknowledgements}
The author acknowledge many fruitful discussions with Hassan Aref and Tomas Bohr. This work was supported by the Materials Research Science and Engineering Center (MRSEC) at Harvard University.
\bibliographystyle{jfm}
\bibliography{KHJ_Slow_Flow_In_Porous_Channels}
\end{document}